\documentclass[aps,prl,twocolumn,superscriptaddress,amsmath,amssymb,nofootinbib,notitlepage,longbibliography
]{revtex4-1}
\usepackage[breaklinks=true]{hyperref}
\hypersetup{
  colorlinks   = true, 
  urlcolor     = blue, 
  linkcolor    = blue, 
  citecolor   =  red 
}

\usepackage{graphicx}
\usepackage[dvipsnames]{xcolor}
\usepackage{cooltooltips} 
\usepackage{bm} 
\usepackage[T1]{fontenc} 
\usepackage{siunitx}

\usepackage[caption=false,subrefformat=simple,labelformat=simple,listofformat=subsimple]{subfig}

\usepackage{dsfont} 
\usepackage[capitalise]{cleveref}

\input{shorthand.sty}

\begin{document}

\title{Hardware-efficient error-correcting codes for large nuclear spins}

\author{Jonathan A. Gross}\email{jonathan.gross@usherbrooke.ca}
\affiliation{Institut quantique and D\'epartement de Physique, Universit\'e de Sherbrooke, Sherbrooke, Qu\'ebec, J1K 2R1, Canada}
\author{Cl\'{e}ment Godfrin}\email{clement.godfrin@usherbrooke.ca}
\affiliation{Institut quantique and D\'epartement de Physique, Universit\'e de Sherbrooke, Sherbrooke, Qu\'ebec, J1K 2R1, Canada}
\author{Alexandre Blais}
\affiliation{Institut quantique and D\'epartement de Physique, Universit\'e de Sherbrooke, Sherbrooke, Qu\'ebec, J1K 2R1, Canada}
\affiliation{Canadian Institute for Advanced Research, Toronto, ON, Canada}
\author{Eva Dupont-Ferrier}\email{eva.dupont-ferrier@usherbrooke.ca}
\affiliation{Institut quantique and D\'epartement de Physique, Universit\'e de Sherbrooke, Sherbrooke, Qu\'ebec, J1K 2R1, Canada}

\date{\today}

\begin{abstract}
Universal quantum computers require a large network of qubits robust against errors.
Recent theoretical and experimental studies on donor nuclear spins in silicon, engineered on  semiconductor platforms compatible with industrial fabrication, show their coherent behavior and potential for scalability.
Here we present a hardware-efficient quantum protocol that corrects phase flips of a nuclear spin using explicit experimentally feasible operations.
We introduce the MAUS encoding (Moment AngUlar System encoding) which uses the large Hilbert space provided by the nuclear spin of the donor to encode the information and employ the electron spin of the donor as an ancilla for error correction. 
Simulations using present-day experimental manipulation fidelities predict significant improvement in logical qubit fidelity over existing spin quantum-error-correction protocols.
These results provides a realizable blueprint for a corrected spin-based qubit.
\end{abstract}

\maketitle

We reach the summit of universal, fault-tolerant quantum computation one step at a time.
Creating a logical qubit that use error correction to outperform bare physical qubits is a key step in this journey.
The community has dubbed this achievement ``beating the break-even point''~\cite{devoret_superconducting_2013}.
So far, the greatest progress toward this goal has been made in superconducting platforms by focusing on correcting the dominant relaxation errors caused by photon loss~\cite{ofek_extending_2016,hu_quantum_2019}.

Dopants in silicon are another very promising platform for quantum information processing~\cite{Morello2020donor}.
Their nuclear spins show record coherence times and single-gate fidelity for a solid-state platform~\cite{saeedi2013room, muhonen2015quantifying}.
Their electron spins allow QND  measurements of the nuclear spins~\cite{pla2013high} and potentially facilitate long-distance coupling of two nuclear spins via the charge degree of freedom~\cite{tosi2017silicon}. 
Recently, two-qubit gates based on exchange coupling were implemented on two-dopant systems~\cite{he2019two, mkadzik2020conditional}.
Additionally, dopants are compact compared to superconducting qubits and can be embedded in devices manufactured in a CMOS foundry~\cite{dupont2013coherent}, allowing them to benefit from the decades of development of the microelectronic industry to integrate them in large-scale platforms based on silicon transistor technology.

Unlike in superconducting circuits, relaxation errors are negligible in donor nuclear spins.
Instead, the relevant errors are dominated by dephasing, as evidenced by the separation between relaxation time $T_1=\SI{65}{s}$ and coherence time $T_2=\SI{60}{ms}$ reported in \cite{pla2013high} and the $T_2=\SI{1.75}{s}$ reported in \cite{muhonen2014storing} for dopants embedded in a nanostructure.
Here we propose a hardware-efficient error-correction scheme that exploits this highly biased noise in a similar spirit to bosonic codes and numerically demonstrate its capacity to beat the break-even point with near-term technology.
Because we choose to encode in a large single spin rather than several small spins to protect against dephasing, our proposal, which we christen the MAUS encoding (Moment AngUlar System encoding), is the ``bosonic code'' version of the protocols outlined in Ref.~\cite{layden_efficient_2020}.
In pursuing the goal of scalable dopant-based architecture, using high nuclear spin offers an advantage as it allows for built-in error correction.

\begin{figure}[t]
  \centering
  \includegraphics[width=8.5cm]{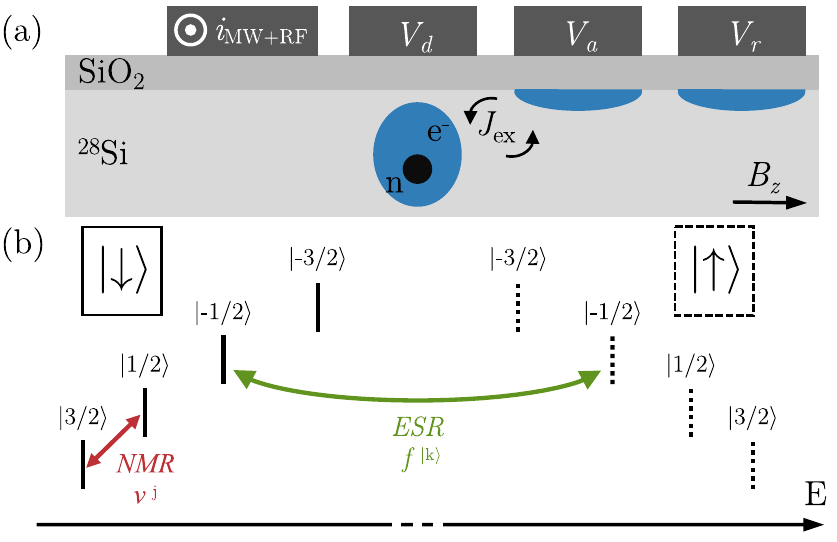}
  \caption{Schematic of the sample for error correcting codes. (a) A donor is implanted in silicon 28. A transmission line is used to apply microwave (radio-frequency resp.) magnetic fields to manipulate the electron (nuclear resp.) spin states. Three gates allow the manipulation of the donor charge state (with potential $V_d$), the ancillary dot (with potential $V_a$) and the readout dot (with potential $V_r$).
  (b) Illustration of the transition manipulations. Nuclear spin energy level diagram for a dopant immersed in $B_z$ magnetic field. For electron spin down (up) state, the state is represented with a solid (dashed) line.
  Two type of manipulation are used in the protocol, nuclear ($\nu^j$) and electron ($f^{\ket{k}}$) spin transition (see text).} 
  \label{fig:qec-system}
\end{figure}

\emph{Describing the system.---}The system we consider---illustrated in \cref{fig:qec-system} (a)---is a single donor implanted in enriched $^ {28}\mathrm{Si}$ whose electron spin $S=1/2$ is coupled to its nuclear spin, $I \ge 3/2 $, via a hyperfine dipole interaction $ A $ and a quadrupole interaction $ Q $.
The magnitude of the nuclear spin is necessary to ensure that the Hilbert space is large enough to detect and recover from at least one error.
The donor electron spin is coupled to the spin of a ancillary accumulation dot via an exchange coupling $J_{\text{ex}}$.
The charge state of the accumulation dot can then be directly measured using a readout dot.
This geometry allows a measurement of the electron spin state of the donor without affecting its charge spatial distribution \cite{nakajima2019quantum,xue2019repetitive}.
Therefore the hyperfine coupling $A$, resulting from a contact interaction, remains constant during the protocol, preventing unwanted nuclear dephasing.
The full spectrum of the spins and the QND aspect of the measurement is discuss in the Supplemental Material~\cite{supp_mat}.
These spins are subjected to a static magnetic field $ B_z $ along the $z$ axis and an oscillating magnetic field $ B_1 $.
This field is generated by an on-chip antenna~\cite{dehollain2012nanoscale} and oscillates in the $x$-$y$ plane at frequency $ f $ to manipulate the spin states.
The Hamiltonian describing this system is
\begin{multline}
H
=
(\gamma_e S_z - \gamma_n I_z) B_z
+ A\,\mathbf{S}\cdot\mathbf{I} + Q I_z^2/4
\\
+ ( \gamma_e S_y - \gamma_n I_y)B_1 \cos (2\pi f t)\,.
\end{multline}

In order to detail the protocol on a minimal system, we focus our discussion on an arsenic donor with nuclear spin $I=3/2$ and depict only the spin transitions of the donor in \cref{fig:qec-system}~(b); however, our protocol generalizes to all donors with a nuclear spin $I \ge 3/2$~\cite{supp_mat}.
For arsenic, $\gamma_e /2 \pi =\SI{28.02}{GHz.T^{-1}}$, $\gamma_n/2 \pi =\SI{7.31}{MHz.T^{-1}}$, $A/2 \pi =\SI{198.35}{MHz}$~\cite{stone2019table}, and $Q/2 \pi$, which depends on the strains at the donor position, is on the order of $\SI{50}{kHz}$~\cite{mourik2018exploring, asaad2020coherent}. Notice that the dipole coupling $A$ can be tuned from the given value down to 0, for donor in its neutral and ionized charge state respectively, with the gate potential $V_d$~\cite{tosi2017silicon}.
To ensure a Zeeman splitting much larger than the electron temperature (necessary for a high-fidelity spin readout) the amplitude of the static field $B_z$ must be on the order of $\SI{1}{T}$.
The typical size of this system is less than $\SI{1}{\micro m^2}$, making it a highly compact corrected systems.
The manipulations applied to the system via magnetic field generated by the current $i_{\mathrm{MW}+\mathrm{RF}}$ [see \cref{fig:qec-system}~(a)] provide both Nuclear Magnetic Resonance (NMR) and Electron Spin Resonance (ESR) [see \cref{fig:qec-system}~(b)].
When applying ESR pulse we drive the electron-spin transition $f^{\ket{k}}$ with the nucleus in state $\ket{k}$, whose frequencies $(\gamma_e B_z + 2 k A) / 2 \pi$ are typically on the order of $\SI{30}{GHz}$.
The theoretical maximum Rabi frequency for this transition is the dipole interaction $A$.
Experimentally, heating effects due to the large current needed to generate the magnetic field limit the Rabi frequency $\Omega$ to a value on the order of $\SI{100}{kHz}$~\cite{muhonen2014storing}. 
When applying NMR pulses, we weakly drive the nuclear-spin transitions $\nu^j$, with $j$ going from 1 to 3 for the transition $\ket{3/2}_z \leftrightarrow \ket{1/2}_z, \ket{1/2}_z \leftrightarrow \ket{-1/2}_z$ and $\ket{-1/2}_z \leftrightarrow \ket{-3/2}_z$ respectively, with frequencies $(\gamma_n B_z + A + (j-2) Q/2) / 2 \pi $.
This pulse, of frequency around $\SI{200}{MHz}$, with an amplitude on the order of $\SI{0.1}{mT}$, induces a Rabi frequency $\Omega$ on the order of $\SI{1}{kHz}$.
This Rabi frequency is sufficiently weak compared to the quadrupole term so as to allow us to address the nuclear transitions individually, otherwise known as the \emph{slow-drive regime}. 
The protocol requires a $I_x$ rotation, which could be applied with a pulse of frequency $(\gamma_n B_z + A)/ 2 \pi $ driven strongly to ensure a Rabi frequency much higher than the quadrupole term.
This would require an oscillating magnetic field larger than $\SI{10}{mT}$, far from the reported experimental limit~\cite{muhonen2015quantifying}.
To implement this $I_x$ rotation we propose using a 3-frequency pulse simultaneously driving all the nuclear-spin transitions. 
This would ensure the $I_x$ rotation stays in the slow-drive regime.
More details are provided on the nuclear spin manipulation in the Supplemental Material~\cite{supp_mat}. 

\emph{Correcting errors.---}The pre\"{e}minence
of $T_2$-type errors in spin systems means that extending the coherence time of logical information is primarily dependent on correcting $I_z$ errors.
Codespaces constructed from extremal eigenstates of an angular-momentum operator in the equatorial plane (such as $I_y$) are ideal for correcting $I_z$ errors.
Expressed in terms of the raising and lowering operators $I_\pm^{(y)}$ for the $y$ component of angular momentum, $I_z=(I_+^{(y)}+I_-^{(y)})/2$, so the effect of an $I_z$ error on the extremal eigenstates of $I_y$ is to decrement the magnitude of the $I_y$ eigenvalues: $\tightket{\pm I}_y\mapsto\tightket{\pm(I-1)}_y$.
As long as $I\geq3/2$, such an error is exactly correctable.
In fact, for $I=(2p+1)/2$ it is possible to correct up to $p$ consecutive $I_z$ errors, analogous to a code of distance $d=2p+1$~\cite[Sec.~2.2]{gottesman_introduction_2009}.
Since extremal eigenstates of angular-momentum operators are spin coherent states, these MAUS codespaces are analogous to the cat codes constructed from coherent states in harmonic oscillators~\cite{cochrane_macroscopically_1999,ralph_loss-tolerant_2005,mirrahimi_dynamically_2014,Puri2017,lescanne_exponential_2020,grimm_stabilization_2020}.

Correction of dephasing errors in oscillators has been discussed autonomously with cat codes~\cite{mirrahimi_dynamically_2014} and in analogy with spin coherent states in binomial
codes~\cite[Sec.~VI.~C.]{albert_performance_2018}.
Native implementation in a spin system, however, is advantageous for at least two reasons.
First, in spin systems, the primary source of error is in fact physical dephasing, unlike in harmonic oscillators where the primary loss channel is usually photon loss.
Second, the measurements and recovery operations required are much more natural in spin systems than the analogous operations in an oscillator.

\begin{figure}[t]
  \centering
  \includegraphics[width=8.5cm]{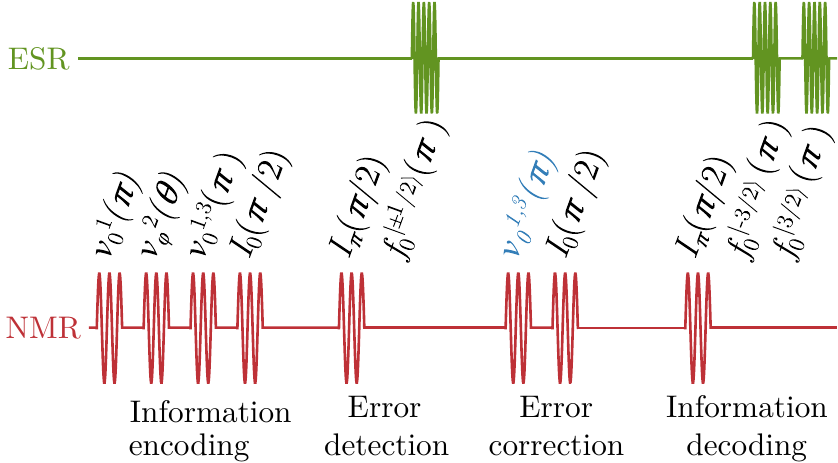}
  \caption{Quantum error correction protocol. Pulse sequence applied to electron (ESR) and nuclear (NMR) spins:  $\nu_{\phi}^{j}(\theta)$ is a rotation of the $j^{\mathrm{th}}$ transitions of the nuclear spin of angle $\theta$ due to a pulse of phase $\phi$. $I_\phi(\theta)$ is a rotation of the nuclear spin of angle $\theta$ due to a pulse of phase $\phi$. $f_{\phi}^{\ket{k}}(\theta)$ is a rotation of angle $\theta$ of the electron-spin transition of nuclear spin $\ket{k}_z$. The  evolution $\nu_0^{1,3} (\pi)$ (displayed in blue) is only applied if an error is detected.}
  \label{fig:qec-seq}
\end{figure}

\begin{figure}[ht]
  \centering
  \includegraphics[width=\columnwidth]{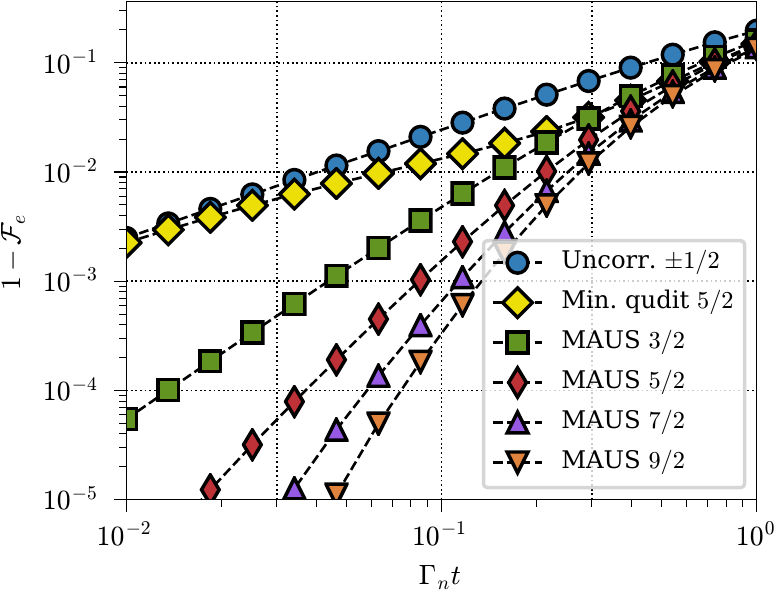}
  \caption{
   Ideal performance of MAUS codes for angular-momentum-eigenbasis-based quantum error correction of dephasing noise, as measured by entanglement fidelity with respect to the noise strength expressed as the product of the nuclear dephasing rate $\Gamma_n$ and the time $t$ between correction cycles.
    The schemes in all spins larger than 1/2 outperform the break even point of encoding the qubit in neighboring $I_z$ angular-momentum eigenstates, which is the encoding that obtains the best fidelity when no error correction is allowed.
    The larger spins exhibit steeper slopes, which illustrates the power of being able to correct random $I_z$ rotations to higher orders.
    The smallest of the minimal qudit codes~\cite{pirandola_minimal_2008} exists in spin 5/2.
    This code does not exactly correct small rotations, so its optimal recovery fidelity~\cite{audenaert_optimizing_2002} exhibits scaling equivalent to a bare qubit in the small-noise limit, only becoming comparable to the MAUS codes when the time between correction cycles is of the order of $T_2\sim1/\Gamma_n$.
  }
  \label{fig:maus-code-performance}
\end{figure}

\begin{figure}[ht]
  \centering
  \includegraphics[width=\columnwidth]{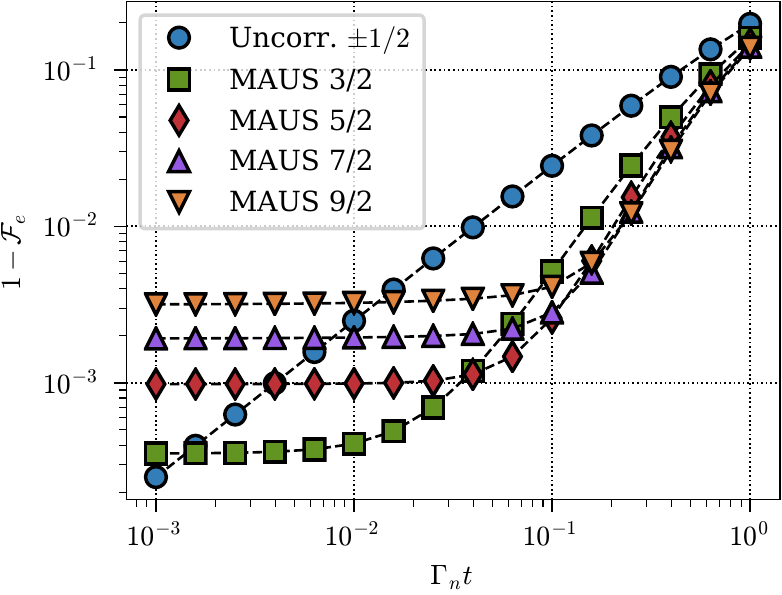}
  \caption{Effect of imperfect detection and recovery operations.
  For relevant nuclear and electron dephasing rates $\Gamma_n=\SI{e-4}\Omega$ and $\Gamma_e=\SI{5e-4}\Omega$ (see text) using the explicit correction protocol described in the text the codes still beat the break-even point for reasonably frequent error correction.
  For the spin-3/2 code we obtain the optimal ratio with respect to the uncorrected $\pm1/2$ encoding when waiting time $t\approx3\times10^{-2}\Gamma_n^{-1}$ between correction cycles.
  A correction cycle involves between two and three $\pi$ rotations between select levels of the nuclear and electron spins, so time required to perform the correction will be at most $3\pi\,\Omega^{-1}$.
  At the noise strength we use in this simulation, this amounts to a ratio between correction time and waiting time of $\pi\times10^{-2}$.
  Going to higher spins at these noise values can bring marginal gains for longer waiting times, but we see the advantage is quickly erased by the more error-prone correction protocol when increasing the repetition rate.
  }
  \label{fig:noisy-maus-correct}
\end{figure}

\begin{figure*}[ht]
  \centering
  \subfloat[][]{
    \includegraphics[width=\columnwidth]{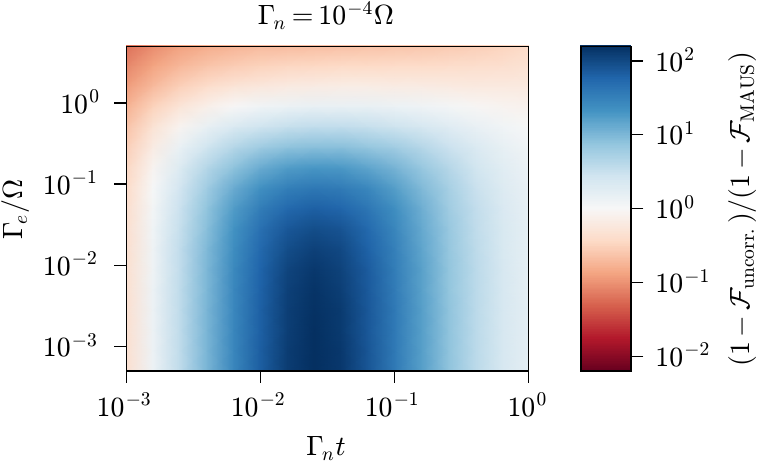}
    \label{fig:vary-electron-dephasing}
  }
  \subfloat[][]{
    \includegraphics[width=\columnwidth]{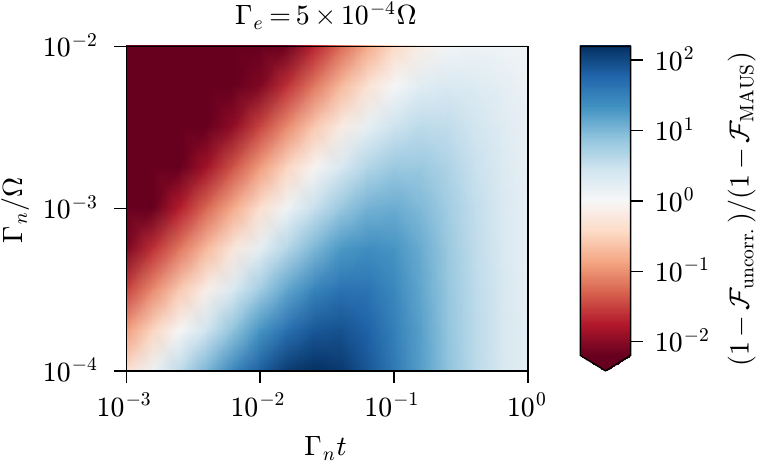}
    \label{fig:vary-nucleus-dephasing}
  }
  \caption{Comparison of our spin-3/2 code to break-even performance over the parameter space.
  The blue regions are where we beat the break-even point.
  Specifically, these regions mark where infidelity of the bare uncorrected qubit (uncorr.) is worse (larger) than the infidelity of the corrected qubit.
  The relevant parameters which we vary are the nuclear dephasing rate $\Gamma_n$ and the electron dephasing rate $\Gamma_e$ (both relative to the Rabi frequency $\Omega$ at which we are able to drive transitions), and the wait time $t$ between correction cycles (expressed relative to the nuclear dephasing rate $\Gamma_n$).
  The smallest value for the electron dephasing rate plotted in (a) is consistent with current experiments, indicating these experiments exceed break-even requirements on readout noise by several orders of magnitude.
  The nuclear dephasing rate for current experiments is also far below the break-even requirement for realistic correction-repetition rates.}
  \label{fig:break-even-regions}
\end{figure*}

\emph{Describing the protocol.---}For the sake of simplicity we illustrate the working principle of our error correction protocol on a 3/2 nuclear spin system (see \cref{fig:qec-seq}).
The same protocol for error correction can be applied to any higher nuclear spin (see \cref{fig:maus-code-performance} and \cref{fig:noisy-maus-correct}) resulting in different performance.
Full protocol for higher nuclear spin is detailed in the Supplemental Material~\cite{supp_mat}.

\textit{Encoding} a logical state starts with preparing the nucleus in the ground state $\ket{3/2}_z$.
To create a superposition of computational-basis states $\ketL{0}=\tightket{3/2}_y$ and $\ketL{1}=\tightket{-3/2}_y$, one performs a $\pi$ pulse to transfer the population from $\ket{3/2}_z$ to $\ket{1/2}_z$, from which one creates the desired logical superposition by manipulating the $\ket{\pm1/2}_z$ subspace.
Parallel $\pi$ pulses transfer the populations from $\ket{\pm1/2}_z$ to $\ket{\pm3/2}_z$.
The maximum time needed for this information-encoding step is that required for a $3 \pi$ nuclear-spin pulse, typically $\SI{1}{ms}$, much shorter than $T_1$ and $T_2$ as are all pulses lengths used for manipulations.
At this point, the information is encoded in the $I_z$ basis, where it is vulnerable to dephasing from the fluctuating magnetic field.
To transfer the information to the $I_y$ basis, where it will be protected from this fluctuating field, one finally applies a linear drive to rotate the entire spin by an angle of $\pi/2$ about the $x$ axis, completing the encoding in the  subspace that protects against dephasing errors.
This encoding procedure is illustrated in the left part of \cref{fig:qec-seq}.

\textit{Error detection} is performed by reversing the $\pi/2$ pulse about the $x$ axis and driving transitions that flip the electron spin when the nuclear spin is in the states $\ket{\pm1/2}_z$.
Notice that the waiting time between two error-detection pulses must be an integer multiple of $4 \pi/Q$, on the order of $\SI{0.2}{ms}$, to remove the evolution induced by the quadrupole term. 
Subsequent measurement of the electron effects a measurement projecting the nuclear spin into the subspace spanned by $\ket{\pm3/2}_z$ (signalling no error) or the subspace spanned by $\ket{\pm1/2}_z$ (signalling an $I_z$ error).
One measures the electron spin of the donor by applying a $\pi$ pulse on the ancillary dot electron only resonant if the donor electron is in its exited state. 
By adjusting the gate potential $V_a$ of the ancillary dot, one can set its chemical potential such that the electron on the ancillary dot can only tunnel to the readout dot when it is in its excited state \cite{elzerman2004single}.
By measuring the presence or absence of such tunneling, one completes the measurement chain, resulting in a QND measurement of the nuclear spin that does not affect the charge state of the donor. 
More details about this error-detection protocol are given in the Supplemental Material~\cite{supp_mat}.
The typical time needed for the electron-spin manipulation is $\SI{0.1}{ms}$. 
Adding this to the QND readout time, on the order of $\SI{0.5}{ms}$ \cite{xue2019repetitive}, gives an error-detection time of $\SI{0.6}{ms}$. 
This is much shorter than the electron-spin lifetime in a static magnetic field of $\SI{1}{T}$, which is on the order of $10^3$ seconds \cite{tracy2013electron}.

During the detection protocol, in the case of an error we allow the electron to remain in its flipped state.
Because of the hyperfine coupling, depending on whether the number of errors during the full protocol is even or odd the $\ketL{0}$ state of the qubit will be encoded in the nuclear-spin state $\ket{3/2}_y$ or $\ket{-3/2}_y$, respectively, which we must keep track of in order to correctly interpret the outcomes of decoding measurements.

\textit{The recovery operation}, to be performed if an error is detected, is simply parallel $\pi$ pulses to transfer population from $\ket{1/2}_z\rightarrow\ket{3/2}_z$ and $\ket{-1/2}_z\rightarrow\ket{-3/2}_z$.
The typical duration of this pulse is $\SI{0.5}{ms}$.
In the event that no error is detected, no recovery operation is needed.
Finally, a $\pi/2$ rotation about the $x$ axis is performed again to return the information to the protected subspace.

\textit{Information decoding} is a $\pi/2$ pulse about the $x$ axis follow by two ESR pulses separated in time.
If the electron spin flip occurs during the first (second) ESR pulse we know that the nuclear spin is in the $\ket{-3/2}_z$ ($\ket{3/2}_z$) state. If the electron spin fails to flip during either of the ESR pulses, this indicates an error has occurred, though a faithful measurement result could still be obtained by using ESR pulses to flip the electron spin conditioned on the nucleus being in the states $\ket{\pm1/2}_z$.

\emph{Noise Effect.---}Detection and recovery operations are necessarily noisy, which implies that the correction step should not be repeated too frequently.
\Cref{fig:noisy-maus-correct} depicts how a code executed with these imperfections can pass from beating the break-even point to performing worse than a bare qubit if the time $t$ between correction cycles is made too short.
Additionally, the fact that detection and recovery are performed while the information is encoded in the $I_z$ basis means that the information is exposed to uncorrectable errors during the procedure, so as usual it is important to perform these operations as quickly as possible.
Even so, this procedure provides a means of extending the lifetime of logical quantum information by storing it predominantly in the protected subspace (where the noise does not corrupt the information).

The baseline against which we compare the performance of our protocol is the uncorrected encoding of logical information in the subspace spanned by $\ket{\pm1/2}$.
This is a subspace that experiences minimal dephasing, and so provides the best storage of logical information in the absence of any error correction. 
For optimal performance, one would use tools from optimal control---such as GRAPE \cite{machnes2011comparing}---to craft optimal pulse sequences for achieving the desired operations.

\emph{Quantifying performance.---}To quantify the performance of these codes and the proposed correction protocol, we perform a series of numerical experiments, using entanglement fidelity/infidelity of the corrected dephasing channel with the ideal identity channel as our figure of merit.
The dephasing channel we use is generated by the master equation
\begin{align}
    \label{eqn:dephasing-lindblad}
    \dot{\rho}
    &=
    \Gamma_n\mathcal{D}[I_z]\rho
    +\tfrac{1}{2}\Gamma_e\mathcal{D}[\sigma_z]\rho\,,
\end{align}
where $\Gamma_n$ and $\Gamma_e$ quantify the dephasing rates of the nuclear and the electron spins, respectively.
\Cref{fig:maus-code-performance} illustrates the performance of the spin-$3/2$ code and the analogous higher-spin codes in an idealized setting where the detection and recovery operations are performed perfectly and instantaneously.
The size of the nuclear spin determines how many $I_z$ dephasing errors can be successfully recovered, so in the limit of small dephasing $\Gamma_nt\ll1$ the infidelity scales as a power of the dephasing that grows with the size of the nuclear spin.
The most analogous existing codes are called \emph{minimal qudit codes}~\cite{pirandola_minimal_2008} and require a nucleus of at least spin 5/2.
As \cref{fig:maus-code-performance} reveals, even our smallest spin-3/2 code outperforms the spin-5/2 minimal qudit code for all but exceedingly large dephasing.

\Cref{fig:noisy-maus-correct} illustrates the fidelities we can expect from this error-correction procedure in the more realistic setting of noisy detection and recovery.
As there is some ambiguity when modeling as to where to locate the imperfections, we choose to concentrate all the noise in the environmental dephasing and leave the control Hamiltonians ideal.
Taking the Rabi frequencies for the transitions between neighboring nuclear and electron spin states to be $\Omega$, we consider nuclear dephasing rates $\Gamma_n$ ranging from  $\SI{e-2}\Omega$ to $\SI{e-4}\Omega$ (for donor charge state going from neutral to ionized, respectively) and electron dephasing rates of at least $\Gamma_e=\SI{5e-4}\Omega$.
These parameter values reflect the greater susceptibility of the electron spin to environmental noise, and result in a nuclear-spin $\pi$-pulse fidelity going from $99\%$ to $99.99\%$ and an electron-spin $\pi$-pulse fidelity of at most $99.95\%$, which is consistent with experimentally measured values reported in the literature \cite{muhonen2015quantifying}.
For nuclear dephasing rate $\Gamma_n=\SI{e-4}\Omega$ one can beat the break-even point with repetition rates less than $10^3\Gamma_n$, giving several orders of magnitude over which we can significantly reduce errors.
We find as much as an order-of-magnitude reduction of entanglement infidelity at the optimal rate $30\Gamma_n$ before hitting a floor when the correction procedure introduces noise more rapidly than the environment.
Larger spins again exhibit more rapid suppression of entanglement infidelity initially as the repetition rate is increased, though the more complicated correction procedure implies a higher noise floor.

\Cref{fig:vary-electron-dephasing,fig:vary-nucleus-dephasing} illustrate parameter regimes in which our protocol beats the break-even point for the spin-3/2 code.
As one can see in \cref{fig:vary-electron-dephasing}, there is a horizontal transition from beating the break-even point to failing to beat that 
point as the electron dephasing rate increases.
This indicates that, for a given nuclear dephasing rate, there is a maximum electron dephasing rate beyond which the error-correction measurements are so noisy as to be useless.
Likewise there is a vertical transition from beating the break-even point to failing to beat the break-even point as the waiting time between correction cycles decreases.
This happens when the time between error-correction cycles is so short that the environmental dephasing is negligible compared to the errors introduced by imperfect detection and recovery operations.
For a fixed electron dephasing rate, as shown in \cref{fig:vary-nucleus-dephasing}, the minimum waiting time above which it is possible to beat the break-even point varies proportionately with the nuclear dephasing rate.
Even for a nuclear dephasing rate of $10^{-2}\Omega$, as one would expect for less favorable neutral donors, one still has a range of correction-cycle-repetition rates spanning an order of magnitude that beats the break-even point.

\emph{Conclusion.---}We have described a hardware-efficient error-correction protocol capable of correcting the most relevant noise in spins.
Our simulations show that this protocol can beat the break-even point given the performance demonstrated in present-day experimental systems.
Our protocol resembles some error-correcting codes for harmonic oscillators, with the advantage that the operations required by the code are native to the spin system.
Recent advances in fast spin readout~\cite{keith2019single, schaal2020fast}, fast QND read-out~\cite{mielke2020nuclear} and coupling between nuclear spins~\cite{hensen2020silicon, tosi2017silicon} will further improve the performance and broaden the applications of our protocol.

This research was undertaken thanks in part to funding from the Canada First Research Excellence Fund and from NSERC.

J.A.G. and C.G. contributed equally to this work.

\bibliography{references.bib}

\section{Supplemental material}

\subsection{Simulations}

The error channel we consider in all simulations is obtained by integrating the Lindblad master equation~\cref{eqn:dephasing-lindblad}.
For~\cref{fig:maus-code-performance}, to illustrate the optimal performance of the codes we consider, the detection and recovery operations are assumed to take place instantaneously.
For the MAUS codes, this means applying a quantum operation that projects the nucleus onto one of the several subspaces spanned by $I_y$ basis vectors of equal magnitude but opposite sign and applies a conditional unitary that maps the resulting subspace back onto the subspace spanned by the extremal $I_y$ basis vectors.
The concatenation of this detection and recovery channel after the noise channel is then applied to half of a maximally entangled pair of nuclei.
The fidelity of the output of this channel with the maximally entangled input state is the entanglement fidelity, which we use as our figure of merit.
To obtain the fidelities we present for the minimal qudit code we solve a semidefinite program~\cite{audenaert_optimizing_2002} to determine the detection and recovery channel that optimizes the entanglement fidelity.

For~\cref{fig:noisy-maus-correct,fig:break-even-regions} we simulate finite-time recovery processes using piecewise-constant control Hamiltonians to drive the transitions of the protocol described in the main text in parallel with the dephasing noise.
All transitions are driven with Rabi frequency $\Omega$ and implemented using the rotating-wave approximation.
This involves first driving the transitions that take the $y$ axis of the spin to the $z$ axis of the spin in order to perform the necessary entangling interactions with the electron.
To generalize the protocol to accommodate spins larger than $3/2$ an iterative procedure is then used where we drive transitions that flip the electron spin conditioned on the projection along the $z$ axis of the nuclear spin having magnitude at most $|I-n|$, for $n$ starting at $1$ and increasing by $1$ each time the electron spin is found to have flipped.
Once the electron spin is found to have not been flipped---or $|I-n|=1/2$---the magnitude of the $z$ projection of the nuclear spin is known, and we drive the transitions that map this subspace back to the subspace spanned by the extremal $I_z$ eigenstates, completing the protocol by finally driving the transitions to take the $z$ axis back to the $y$ axis.
This arrangement of the detection measurements is chosen to minimize the total detection and recovery time, as the most-likely scenarios involve small decreases in the magnitude of the angular-momentum projection along $z$, which are identified by the protocol in fewer measurements than more dramatic magnitude reductions.

\subsection{Coherent manipulations}

In this section, we detail the nuclear and electron spin manipulation Hamiltonians and we discuss the microwave-pulse properties as well as the fidelities we expect given recent experimental results.
A spin subjected to a resonant monochromatic microwave pulse rotates at an angular frequency $\Omega$ around an axis $\mathbf{r} = (\sin \phi , -\cos \phi,0 )$, $\phi$ being the phase of the pulse.
The spin state at any time $\tau$ can be calculated by the equation $\ket{\Psi(\Omega \tau)} = R_{\phi} (\Omega \tau) \ket{\Psi(0)} $.
The rotation operator $R$ depends on the applied pulse.

\subsubsection{Nuclear spin}

The record reported fidelity of a coherent manipulation of a single donor nuclear spin in enriched silicon 28 is $99.99\%$~\cite{muhonen2015quantifying}.
This fidelity is obtained when the donor is in its ionised state. The reported fidelity for the donor in its neutral state is $99\%$~\cite{muhonen2015quantifying}.   
Because we need a non null dipole coupling term $A$ to implement the protocol, we take these two values as extreme case in the calculation of the code fidelity \cref{fig:vary-nucleus-dephasing}.
All NMR manipulations are applied in the slow-manipulation limit ($\Omega_{\nu} \ll Q$), meaning the applied oscillating magnetic field is much smaller than $\SI{0.5}{mT}$. 
Each transition can then be manipulated individually.
In this regime, the generalised rotating-frame approximation is appropriate to describe the dynamics of the nuclear spin~\cite{leuenberger2003grover}.

We first discuss the quasi mono-chromatic regime. In this regime, each nuclear spin is subjected to a maximum of one resonant pulse. Under this approximation, a pulse of frequency $\nu_{pq}$, phase shift $\phi$ and duration $\tau$, resonant with the nuclear spin transition $\ket{p} \leftrightarrow \ket{q} $, induces a Rabi oscillation of angle $\theta= \Omega_{pq} \tau$.
The expression for the rotation unitary is then
\begin{equation}
  R^{\ket{p}, \ket{q}}_{\phi} (\theta) = \exp[i \theta (\sigma_x^{\ket{p}, \ket{q}} \cos \phi + \sigma_y^{\ket{p}, \ket{q}}\sin \phi) /2]\,,
\end{equation}
with $\sigma^{\ket{p},\ket{q}}_{x,y}$ being Pauli operators acting on the subspace spanned by $\ket{p}$ and $\ket{q}$.
As shown in~\cref{fig:qec-seq}, we use this unitary evolution during the information-encoding step ($\pi$ pulse on the first transition, arbitrary pulse on the second transition and simultaneous $\pi$ pulse on the first and the third transition) and the error-correction step (simultaneous $\pi$ pulse on the first and the third transition).
The typical frequency of the microwave pulse sent to perform these manipulations is $\SI{200}{MHz}$. 
For the $I_x$ rotation, the dynamic is different in the sense that a given state can be subjected to two components of the poly-chromatic pulse. We then need to treat the generalised rotating frame Hamiltonian in its more general form. Considering pulse frequency component $\nu_j = \gamma_n B_z + A + (2-j) Q $ of amplitude $B_j$ resonant with the nuclear spin transition $j$, the Hamiltonian of the nuclear spin dynamic is: 
\begin{equation}
H_{G.R.F} = \gamma_n /2  
\begin{pmatrix}
0 & \sqrt{3} B_1 & 0 & 0\\
\sqrt{3} B_1 & 0 & 2 B_2 & 0\\
0 & 2 B_2 & 0 & \sqrt{3} B_3\\
0 & 0 & \sqrt{3} B_3 & 0\\
\end{pmatrix}
\end{equation}

It is then possible to calibrate the magnetic field amplitude of each pulse frequency component to reach the equality $H_{G.R.F}=I_x$. 
This manipulation is used before and after each free evolution.
The typical frequencies of the microwave pulse component sent to perform these manipulations are $\SI{200}{MHz}$.

\subsection{QND read-out}\label{sec:qnd}

A key point of the protocol is the quantum non demolition (QND) read out of the nuclear spin.
Even though \cref{fig:qec-system} only depicts the double dot approach, we discuss two approaches in this section. 
The first is the one implemented in the experiment that gave the state-of-the-art coherent-manipulation fidelity for donors~\cite{muhonen2015quantifying}, the second is a proposal that would induce less dephasing during the error detection step. 

\subsubsection{Single-dot measurement}

To find out if the nuclear spin is in the $\{ \ket{-1/2}, \ket{1/2} \}$ subspace, we apply a polychromatic pulse ($f^{\ket{-1/2}}$ and $f^{\ket{1/2}}$) of amplitude and duration guaranteeing a $\pi$ rotation of the electron spin.
If the nuclear spin is in the subspace $\{ \ket{-3/2} , \ket{3/2} \}$ the pulse won't have any effect on the system.
By contrast, if the nuclear spin is in $\{ \ket{-1/2} , \ket{1/2} \}$ the pulse is resonant and will reverse the electron spin.
This change of electron spin induces an electron transfer back and forth with the lateral dot tuned with a appropriate chemical potential.
This electron transfer with the dot can be measured by a transport measurement through a SET or by rf-reflectometry on the dot.
The electron transfer between the donor and the dot impacts the nuclear spin dynamic.
Indeed, the harmonic $A$ term of the hyperfine coupling comes from a contact interaction between the nuclear spin and the electron spin.
When the electron tunnels back and forth between the donor and the dot, the nuclear spin is no longer subject to the same dynamics.
This induces a loss of information on the phase of the nuclear spin $\delta \phi= A \delta t$, $\delta t$ being the time resolution of the electron position measurement.
The hyperfine interaction term $A$ is of the order of $\SI{e8}{Hz}$ and the fastest charge readout, based on reflectometry measurement \cite{Stehlik2015Fast,Connors2020Rapid}, is of the order of $\SI{e-7}{s}$, giving a phase uncertainty of the order of $\SI{10}{\radian}$.
To reduce this uncertainty down to $10^{-3}$, a value where this dephasing mechanism could be negligible compared with the environmental dephasing process, it is necessary both to reduce the coupling term $A$ and to use a faster charge-readout technique.
A proposal~\cite{tosi2017silicon} shows that by applying a DC electric field on top of the donor, the electron wave function can be delocalised to an interface charge state.
It is then possible to tune the $A$ term from zero to its zero field value.
Notice that the value of $A$ cannot be null in the proposed protocol to keep a sufficiently high energy difference in between the different electron spin transitions.
This decrease of $A$ added to the temporal resolution offered by reflectometry techniques  would lower the phase uncertainty.
However, in order to have a reasonable phase uncertainty, technical developments are necessary, which is why we propose a double-dot system.

\subsubsection{Double-dot measurement}

\begin{figure*}[ht]
  \centering
  \includegraphics[width=\textwidth]{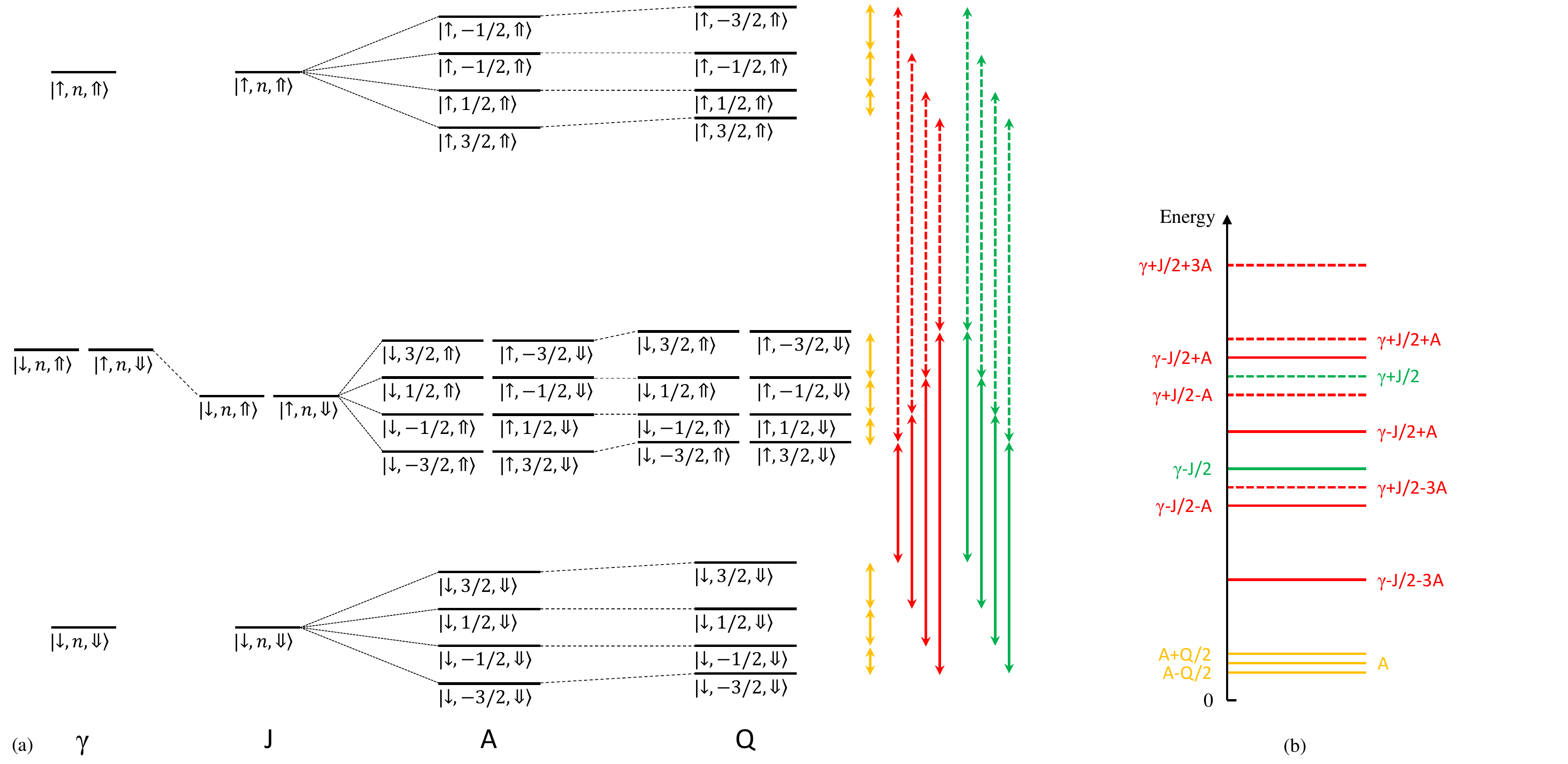}
  \caption{Spin transitions of the donor/double-dot system.
  (a) At each column an interaction is added to display how the degeneracies are lifted.
  On the right (b), the spin transitions are ordered as function of their energies.  The nuclear spin transitions $\nu^j $ are in yellow,  the error detection pulses $f^{\pm 1/2}_0$ are in solid red line ($\gamma-J/2 \pm A$) and the ancillary-dot-electron transition is in solid green line ($ \gamma-\pm J/2 $). This spectrum shows that the transitions of the donor and dot electron spins are at two different energies due to the hyperfine coupling.
  Besides, the dot electron-spin manipulation is conditioned by the electron spin to the donor as needed to perform the QND readout without modifying the charge state of the donor.}
  \label{fig:qec-spectreadout}
\end{figure*}

\Cref{fig:qec-system} (a) illustrates another idea to avoid the dephasing effect due to the modification of the hyperfine coupling during the readout process.
The idea is to use a read-out technique that does not affect the charge state of the donor.
Such a technique has been demonstrated in several different systems \cite{godfrin2017electrical, nakajima2019quantum}, including silicon based quantum dots \cite{xue2019repetitive}.
In this last case, a $\pi$ pulse is applied to the electron spin below $V_a$.
This pulse is resonant with the electron-spin transition only if the donor electron spin is in its excited state.
The electron-spin flip induces a transfer back and forth of the ancillary electron with the SET as for the actual donor spin readout. 
Notice that in the donor system, the degeneracy between the dot and the donor electron spin is lifted by the hyperfine coupling, meaning that no micro-magnets are needed (see \cref{fig:qec-spectreadout}). 
In this protocol, a microwave pulse needs to be added to the sequence of \cref{fig:qec-seq}.
In the error-detection part, after the $ f^{\pm 1/2}_0 (\pi)$ pulse, a $\pi$ pulse is applied to the electron spin located under $V_a$. 
This pulse is resonant only if the donor electron spin has been flipped by the $ f^{\pm 1/2}_0 (\pi)$ pulse.
This exchange coupling may affect the fidelity of the electron-spin manipulation. 
In addition, the charge state of the donor is kept in its neutral sate. 
This avoids the uncertainty of the previous protocol but induces a bigger constant dephasing due to a higher coupling of the nuclear spin to the environment through the hyperfine coupling. 
Indeed, as measured \cite{muhonen2014storing}, the nuclear spin $T_2$ goes from 1.75s in its ionized state to 20.4ms in its neutral state.
However, the hyperfine coupling $A$ can be tuned from its zero-field value down to zero with the gate potential $V_d$~\cite{tosi2017silicon}. An intermediate value of $A$ allows lower dephasing rates while keeping a non-zero value, necessary for the pulse selectivity of the protocol. \Cref{fig:vary-nucleus-dephasing} displays a significant gain of the MAUS code compared with a bare qubit for a range of nuclear-spin dephasing rates going from its neutral to ionized charge-state value.

\subsection{Protocol for higher order spin}
Increasing the size of the nuclear spin increases the complexity of the pulse sequence, especially for the error detection. 
Here we briefly describe the protocol for a $\ket{5/2}$ nuclear spin.
To limit the average number of manipulations required, we first flip the electron spin conditioned on one or two errors having occurred.
Most of the time, no error will be detected at this stage. 
If an error is detected, a second pulse is then sent to distinguish between one and two errors having occurred.
The same logic can be generalized to higher spins, making sure that each binary measurement can unambiguously identify the most likely error at each step. 

\begin{figure*}[ht]
  \centering
  \includegraphics[width=\textwidth]{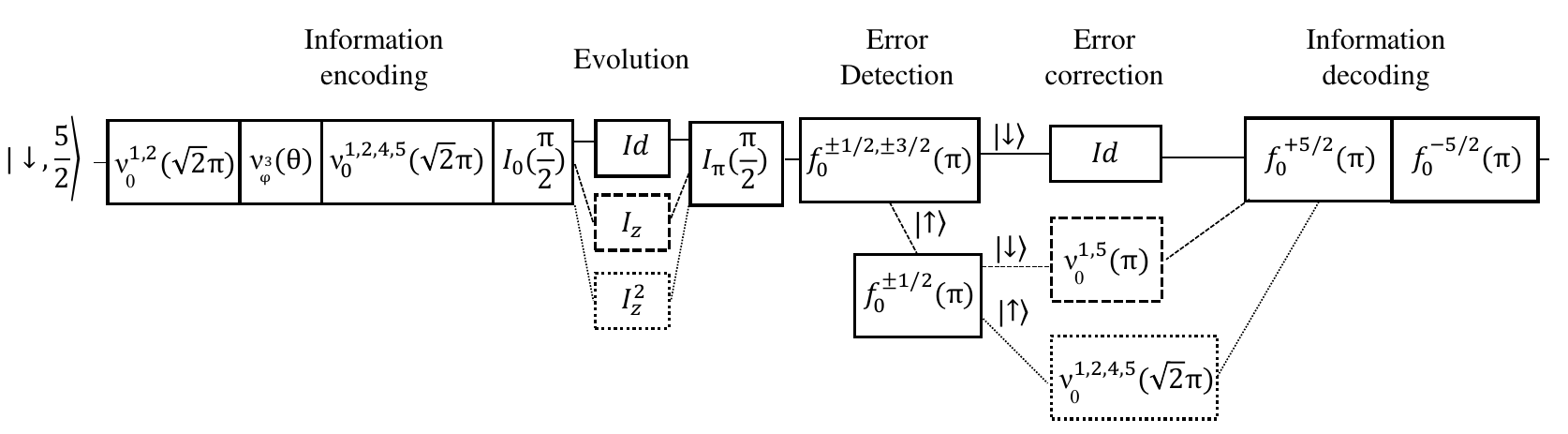}
  \caption{Illustration of the MAUS error correcting protocol applied to a spin 5/2. 
  }
  \label{fig:qec-seqsuping}
\end{figure*}

\end{document}